\documentclass[useAMS,usenatbib]{mn2e}

\def\rms{r_{\rm ms}}
\def\xms{x_{\rm ms}}
\def\dM{\dot{M}}
\def\dMign{\dot{M}_{\rm ign}}

\def\dMtr{\dot{M}_{\rm trap}}

\def\dE{\dot{E}_{\nu\bar\nu}}
\def\Teff{T_{\rm eff}}
\def\Lobs{L_{\rm obs}}

\def\dn{\dot{n}_{\nu\bar\nu}}
\def\dN{\dot{N}_{\nu\bar\nu}}
\def\dd{d}
\def\simlt{\lower.5ex\hbox{$\; \buildrel < \over \sim \;$}}
\def\simgt{\lower.5ex\hbox{$\; \buildrel > \over \sim \;$}}

\usepackage{graphicx}

 \title[Neutrino heating near hyper-accreting black holes]
{Neutrino heating near hyper-accreting black holes}
\author[Ivan Zalamea and Andrei M. Beloborodov]
{Ivan Zalamea\thanks{E-mail: izalamea@phys.columbia.edu} and
Andrei M. Beloborodov\thanks{Also at Astro-Space Center of Lebedev Physical
Institute, Profsojuznaja 84/32, Moscow 117810, Russia} \\
Physics Department and Columbia Astrophysics Laboratory,
Columbia University, New York, NY 10027, USA\\
}
\begin{document}

\date{Accepted ----. Received -------; in original form ----}

\pagerange{\pageref{firstpage}--\pageref{lastpage}} \pubyear{-----}

\maketitle

\label{firstpage}

\begin{abstract}
Hyper-accretion discs around black holes emit copious neutrinos and 
anti-neutrinos. A fraction of the emitted neutrinos convert to 
electron-positron plasma above the disc through the annihilation reaction 
$\nu\bar\nu\rightarrow e^+e^-$. This process may drive relativistic jets 
associated with GRB explosions.
We calculate the efficiency of energy deposition by neutrinos.
Our calculation is fully relativistic and based on a geodesic-tracing
method. We find that the efficiency of 
neutrino heating
is a well-defined function of
(i) accretion rate and (ii) spin of the black hole. It is practically
independent of the details of neutrino transport in the opaque zone
of the disc. The results help identify accretion discs whose
neutrino emission can power GRBs.
\end{abstract}



\section{Introduction}

A plausible model for the central engines of gamma-ray bursts (GRBs)
pictures a transient, hyper-accreting disc formed around a rotating 
black hole (see Beloborodov 2008 for a recent review). 
The disc is a source of copious neutrinos and anti-neutrinos, which 
partially annihilate above the disc 
and turn into $e^\pm$ pairs, $\nu+\bar{\nu}\rightarrow e^-+e^+$. 
This process was proposed as a possible mechanism 
for creating relativistic, $e^\pm$-dominated jets that could power 
observed GRBs (Eichler et al. 1989). However, 
its efficiency
remained unsettled. Its accurate calculation requires a detailed 
relativistic model for the neutrino source --- the accretion disc --- 
as well as tracing the neutrino propagation in the Kerr spacetime of
the black hole.

A detailed relativistic model for GRB discs was completed recently
(Chen \& Beloborodov 2007, hereafter CB07). It describes the disc 
down to its inner edge and gives accurate energy fluxes carried 
away by $\nu$ and $\bar{\nu}$ at all radii $r$. 
In the present work we trace 
the
neutrino trajectories 
and calculate the rate of $\nu\bar{\nu}$ annihilation around the disc. 
Neutrino annihilation was previously calculated in a number of works 
(e.g. Popham, Woosley \& Fryer 1999.; Asano \& Fukuyama 2001; 
Birkl et. al. 2007). 
Our work has three motivations:  (i) A relativistic calculation of 
$\nu\bar{\nu}$ annihilation has never been 
done for a realistic accretion disc around a spinning black hole. Previous 
works either used a toy model for neutrino source (e.g. 
Birkl et al. 2007) or replaced neutrino trajectories by straight lines 
(Popham et. al. 1999).
(ii) The efficiency of $\nu\bar\nu$ annihilation 
strongly depends on the accretion rate $\dot{M}$ and the spin parameter 
$a$ of the black hole. It is desirable to know this dependence and identify 
the range of $\dot{M}$ and $a$ where $\nu\bar\nu$ annihilation can provide 
the observed energy of GRB explosions. 
(iii) In addition 
to
$\nu\bar{\nu}$ annihilation, neutrinos can create $e^\pm$
pairs off magnetic field (Kuznetsov \& Mikheev 1997; Gvozdev \& Ognev 2001). 
The contribution of this process 
to the energy deposition rate 
should be included.

The paper is organized as follows.
Section~\ref{sec:model} describes the setup and method of our calculations.
The results are presented in Section~\ref{sec:results} and summarized 
in Section 4.

\section{Model description}\label{sec:model}

Neutrinos emitted by the disc follow geodesics in Kerr spacetime.
The efficiency $\epsilon$ of their annihilation can be calculated
numerically by tracing the geodesics, evaluating the local energy deposition
rate [erg~s$^{-1}$~cm$^{-3}$] everywhere around the black hole
and then integrating over volume 
outside the disc
to obtain the net 
energy deposition rate $\dE$ (energy at infinity per unit time at infinity).

The neutrino emission and annihilation is concentrated near the black hole,
where the accretion time-scale is short and $\dot{M}$ may be assumed 
to be quasi-steady. Then $\dE$ depends on four parameters that
specify the steady disc model: accretion rate $\dM$, viscosity parameter
$\alpha$, mass of the black hole $M$, and spin of the black hole $a$.

\subsection{Neutrino source: disc model}

As the disc matter spirals into the black hole, it is viscously heated: 
the gravitational energy is converted to heat. Part of the heat is lost 
to neutrino emission, and part is stored in the disc and
distributed between nuclear matter, radiation, and $e^\pm$ pairs, 
in perfect thermodynamic and nuclear statistical equilibrium.
The equilibrium microphysics is determined by only three parameters:
temperature $T$, baryon mass density $\rho$, and electron fraction $Y_e$
(equal to the charged nucleon fraction). Other parameters ---
e.g. the electron chemical potential $\mu_e$ and density of $e^\pm$ pairs
$n_\pm$ --- are derived from $T$, $\rho$ and $Y_e$. 

The neutrino emission peaks in the inner region of the disc.
The far dominant emission mechanism is the $e^-$/$e^+$ capture onto
protons/neutrons,
\begin{equation}
\label{eq:reactions}
  e^-+p\rightarrow n+\nu_e, \qquad e^++n\rightarrow p+\bar{\nu}_e.
\end{equation}
The neutrino-cooled discs are nearly in $\beta$-equilibrium: the rates
of the two reactions in equation~(\ref{eq:reactions}) are practically 
equal, which determines the value of $Y_e(\rho,T)$ (Imshennik et al. 1967; 
Beloborodov 2003).
In principle, all three flavors of neutrinos could be emitted from the disc, 
but only electron neutrinos need to be considered; the emission rates for 
muon and tau neutrinos are negligible (CB07). 

A detailed model for neutrino-cooled relativistic discs was developed in 
CB07, and we use their model in our calculations.
The approximate hydrodynamic disc equations are solved with the 
vertically-integrated $\alpha$ prescription. The equations include 
radial transport of heat and lepton number. Local microphysics is 
treated exactly: nuclear composition, 
electron degeneracy, neutrino emissivity and opacity etc., using the 
equilibrium distribution functions for all species except neutrinos. 
Neutrinos are modeled separately in the opaque and transparent zones of 
the disc, matching at the transition between the two zones.

The disc model provides the vertically averaged $T$, $\rho$, $Y_e$, $\mu_e$ 
(which are approximately equal to their values in the midplane $\theta=\pi/2$)
and the half-thickness of the disc $H$.
The $\nu$ and $\bar\nu$ spectra emitted from the neutrino-transparent
(optically thin) zone of the disc are given by
\begin{eqnarray}
\label{eq:fneutrinos}
\nonumber
   f(E)   & \equiv & h^3\frac{dN_\nu}{\textrm{d}^3x\,\textrm{d}^3 p} \\
           &  =  & \frac{\lambda^3KH\rho Y_e}{\pi m_pc} \,
   \frac{(E+q)\sqrt{(E+q)^2-1}}{e^{(E+q-\mu_e)/\Theta}+1}
\end{eqnarray} 
for neutrinos ($E>0$), and 
\begin{eqnarray}
\label{eq:fantineutrinos}
\nonumber
   f^\prime(E) & \equiv &  h^3\frac{dN_{\bar\nu}}{\textrm{d}^3
                         x\,\textrm{d}^3p}\\
                 &  =  & \frac{\lambda^3 K H\rho (1-Y_e)}{\pi m_p c}
  \frac{(E-q)\sqrt{(E-q)^2-1}}{e^{(E-q+\mu_e)/\Theta}+1} 
\end{eqnarray}
for anti-neutrinos ($E>q+1$). 
Here $h=2\pi\hbar$, $\lambda=h/m_ec$, $q=(m_n-m_p)/m_e=2.53$, 
$K=6.5\times 10^{-4}s^{-1}$, $\Theta=kT/m_{e}c^{2}$,
$\mu_e$ is electron chemical potential in units of $m_ec^2$
and $E$ is neutrino/anti-neutrino energy in units of $m_ec^2$. 

The spectra emerging from the opaque zone are controlled by neutrino 
transport through the disc, which cannot be reliably calculated --- 
it depends on the unknown vertical distribution of viscous heating.
Fortunately, the $\nu\bar\nu$ annihilation rate depends
only on the \emph{energy fluxes} 
$F_\nu$ and $F_{\bar\nu}$ from the disc surface, which are insensitive 
to the neutrino-transport details. 
This fact has a simple analytical explanation (Beloborodov 2008).
It follows from the proportionality 
$\sigma_{\nu\bar\nu}\propto E_\nu E_{\bar\nu}$, where 
$\sigma_{\nu\bar\nu}$ is the cross-section for annihilation for 
neutrino and anti-neutrinos with energies $E_\nu$ and $E_{\bar\nu}$.

To demonstrate that the rate of $\nu\bar\nu$ annihilation depends on
$F_\nu$ and $F_{\bar\nu}$ but not on the exact shapes of $\nu$ and 
$\bar\nu$ spectra, we consider below two extreme models A and B, 
and they give practically the same annihilation rates:

 {\bf Model~A:} Neutrinos 
     and anti-neutrinos are emitted with the same spectra as found
     \emph{inside} the disc. The spectra are normalized so that 
the emerging  emission carries away the known energy fluxes $F_{\nu}$ 
and $F_{\bar\nu}$. 
In the opaque region, 
the spectra of $\nu$ and $\bar\nu$ are 
  described by Fermi-Dirac distributions. The temperature and chemical 
  potential for $\nu$ and $\bar\nu$ inside the disc are obtained from the 
  numerical models of CB07.

{\bf Model~B:} Neutrinos
     and anti-neutrinos are emitted with a thermal spectrum with zero chemical
     potential and temperature $T=T_{\rm eff}$, where $T_{\rm eff}$ is the 
     effective surface temperature
defined by $(7/8)\sigma T_{\rm eff}^4=F_\nu+F_{\bar\nu}$
  ($\sigma$ is the Stefan-Boltzmann constant and the coefficient 7/8 
   takes into account the difference between statistics of photons 
   and $\nu$, $\bar\nu$).

When the disc is efficiently cooled (neutrino energy flux almost balances 
viscous heating), $\Teff$ is given by the standard thin-disc model 
of Page \& Thorne 
  (1974): 
$\Teff\approx\Teff^{\rm st}$.
This regime occurs in a broad range of accretion rates $\dMign<\dM<\dMtr$ 
(CB07). If $\dM<\dMign$, the disc temperature is not high enough to ignite 
the neutrino emitting reactions. If $\dM>\dMtr$, the emitted neutrinos 
become trapped in the disc and advected into the black hole. 
Our third (simplest) model for the neutrino source is defined as follows.

{\bf Model~C:} Neutrinos and anti-neutrinos
are emitted with a thermal 
spectrum that has zero chemical potential and the following temperature,
\begin{eqnarray}\nonumber
   \label{eq:C}
    \Teff(\dM,r)
    =
    \Teff^{\rm st}(\dMign,r)\qquad\qquad\qquad\qquad\\\
   \qquad\quad \times 
       \left\{ \begin{array}{ll}
      0                 &  \dM<\dMign \\
  (\dM/\dMign)^{1/4}    &  \dMign<\dM<\dMtr \\
  (\dMtr/\dMign)^{1/4}  &  \dM>\dMtr \\
               \end{array}
       \right.
\end{eqnarray}
This model does not even require the calculation of the disc structure 
as $T_{\rm eff}^{\rm st}$ is a known analytical function of $r$ 
  (Page \& Thorne 1974).
As we show below, this simplest model gives remarkably accurate result 
for $\dE$. 

The characteristic accretion rates $\dMign$ and $\dMtr$ depend 
on the viscosity parameter $\alpha$. They were calculated in CB07,
and the numerical results are well approximated by the following 
formulae,
\begin{eqnarray}\label{eq:pw}
  \dMign=K_{\rm{ign}}\left( \frac{\alpha}{0.1} \right)^{5/3},\qquad
  \dMtr=K_{\rm{trap}}\left( \frac{\alpha}{0.1} \right)^{1/3}.
\end{eqnarray}
The coefficients $K_{\rm{ign}}$ and $K_{\rm{trap}}$ are functions of
the black-hole spin $a$. For $a=0$, 
$K_{\rm{ign}}=0.071M_{\odot}s^{-1}$ and 
$K_{\rm{trap}}=9.3 M_{\odot}s^{-1}$. For $a=0.95$, 
$K_{\rm{ign}}=0.021M_{\odot}s^{-1}$ and 
$K_{\rm{trap}}=1.8 M_{\odot}s^{-1}$.

In all three models A, B and C the neutrino emission is assumed
to be isotropic in the local rest frame of the disc (which is in Keplerian 
rotation around the black hole).

\subsection{Neutrino transport}

To evaluate the $\nu\bar\nu$ annihilation rate at a given point we need
to know the local $\nu$ and $\bar\nu$ distribution functions. 
They can be obtained 
using the known neutrino distribution functions at the surface of the disc.
To a first approximation, neutrinos 
obey the collisionless Boltzmann equation, because most of them do not 
participate in any interactions (and eventually escape to 
     infinity or get captured into the black hole).
The Boltzmann equation in curved spacetime has the same form as in 
flat spacetime. It states that the phase-space density (or occupation 
number) of neutrinos remains constant along their trajectories,
\begin{eqnarray}\label{eq:Boltzman}
  \frac{d f(x^{\mu}(\lambda),p^{\mu}(\lambda))}{d\lambda}=0.
\end{eqnarray} 
Here $x^{\mu}(\lambda)$ is a parameterized worldline for a $\nu$ or $\bar\nu$.
The neutrinos emitted by the disc have huge energies 
compared to their rest mass and we treat them as massless particles 
propagating along null geodesics in Kerr spacetime. 
We use Boyer-Lindquist coordinates $x^\alpha=(t,\phi,r,\theta)$, 
where the Kerr metric has the form,
\begin{equation} 
  ds^2=g_{tt}dt^2+2g_{t\phi}dtd\phi +g_{\phi\phi}d\phi^2
      +g_{rr}dr^2+g_{\theta\theta}d\theta^2.
\end{equation}
The metric tensor $g_{\alpha\beta}$ is specified by two parameters:
$r_g=2GM/c^2$ and the spin parameter $|a|<1$; 
it is given in e.g. Chandrasekhar (1998).
 
Equation~(\ref{eq:Boltzman}) is covariant and its solution takes into 
account Doppler and gravitational redshifts. The null geodesics in 
Boyer-Lindquist coordinates are described by 
 the 
known ordinary differential 
equations of first order (e.g. Chandrasekhar 1998) which we solve numerically.

\begin{figure}
\begin{center}
\includegraphics[width=2.6in]{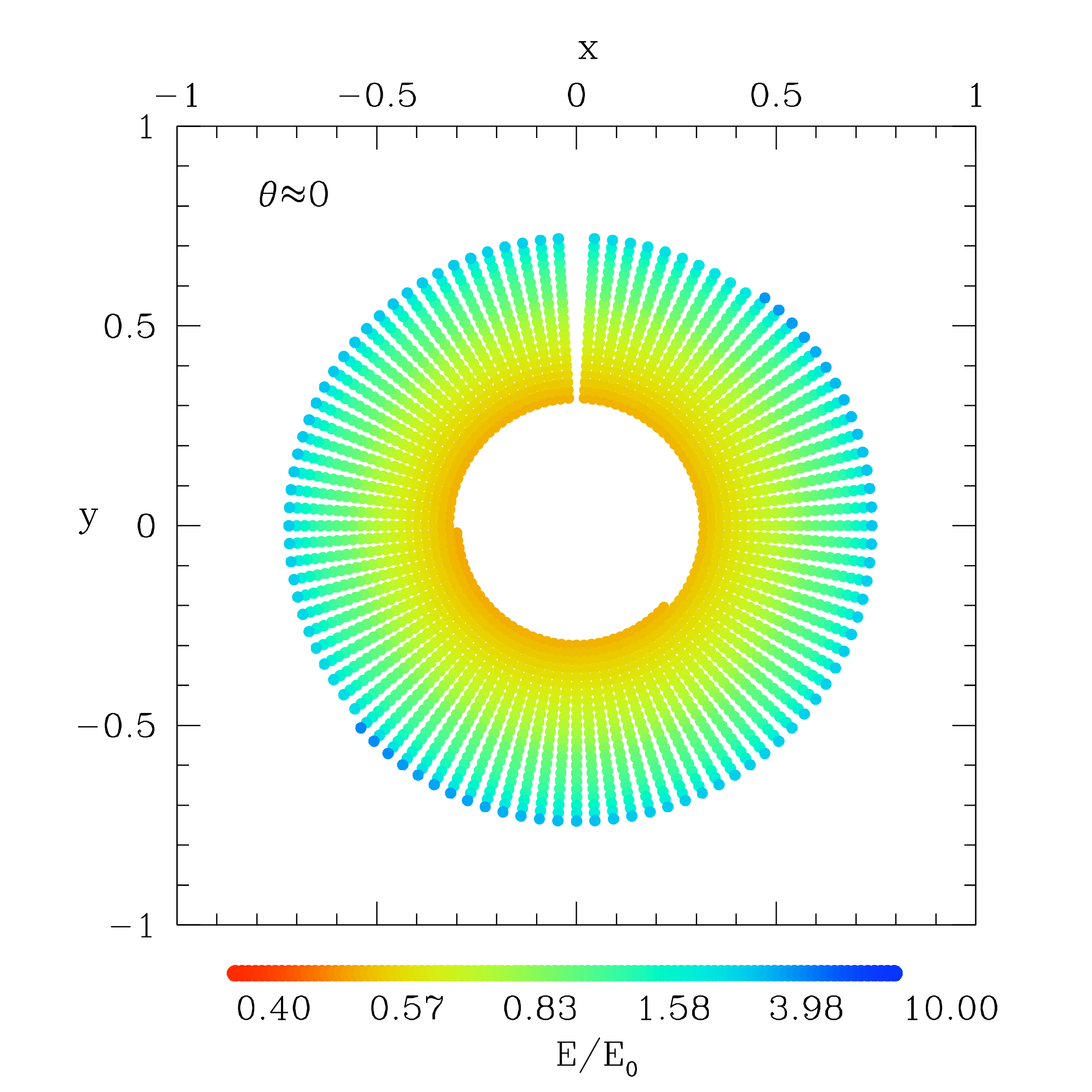}
\includegraphics[width=2.6in]{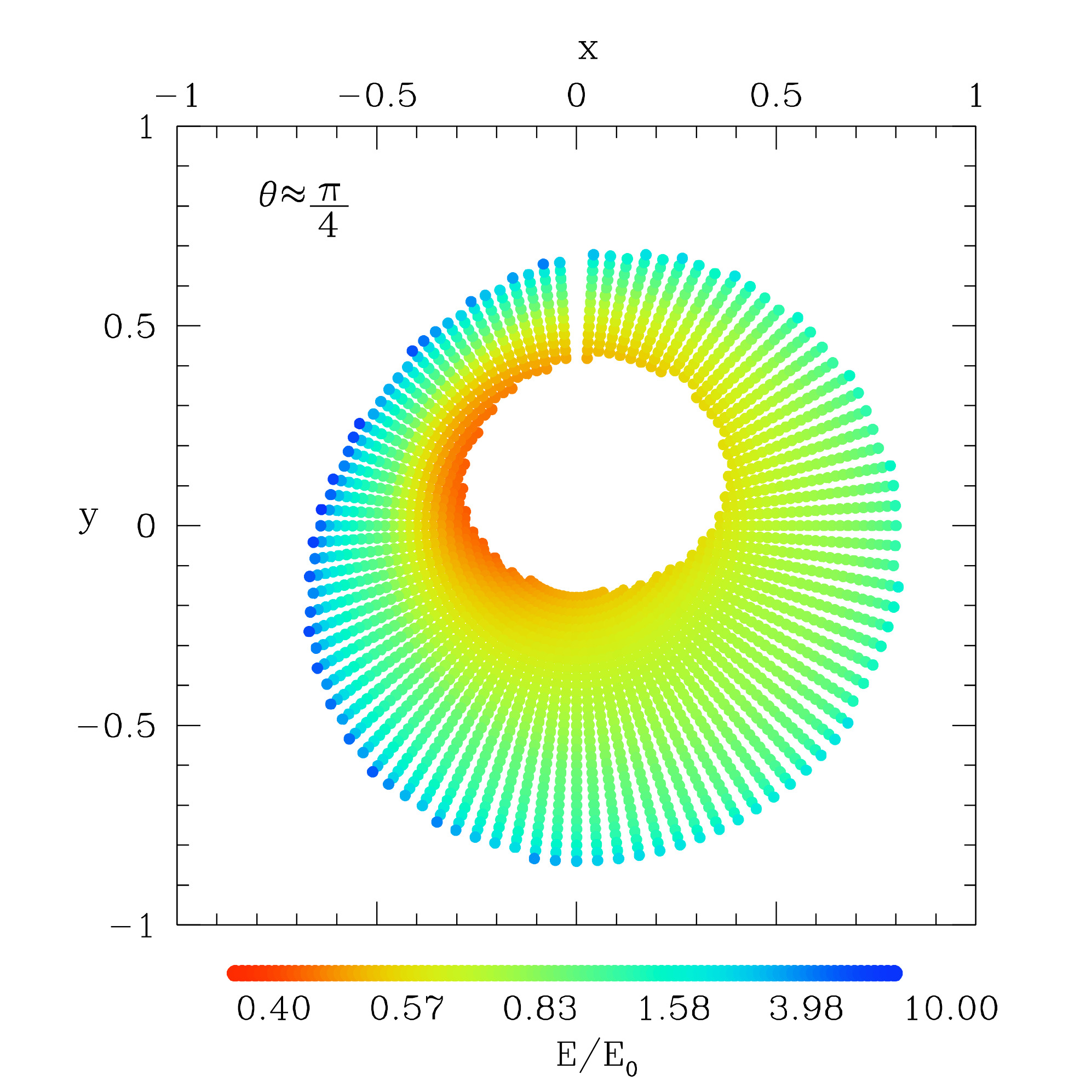}
\includegraphics[width=2.6in]{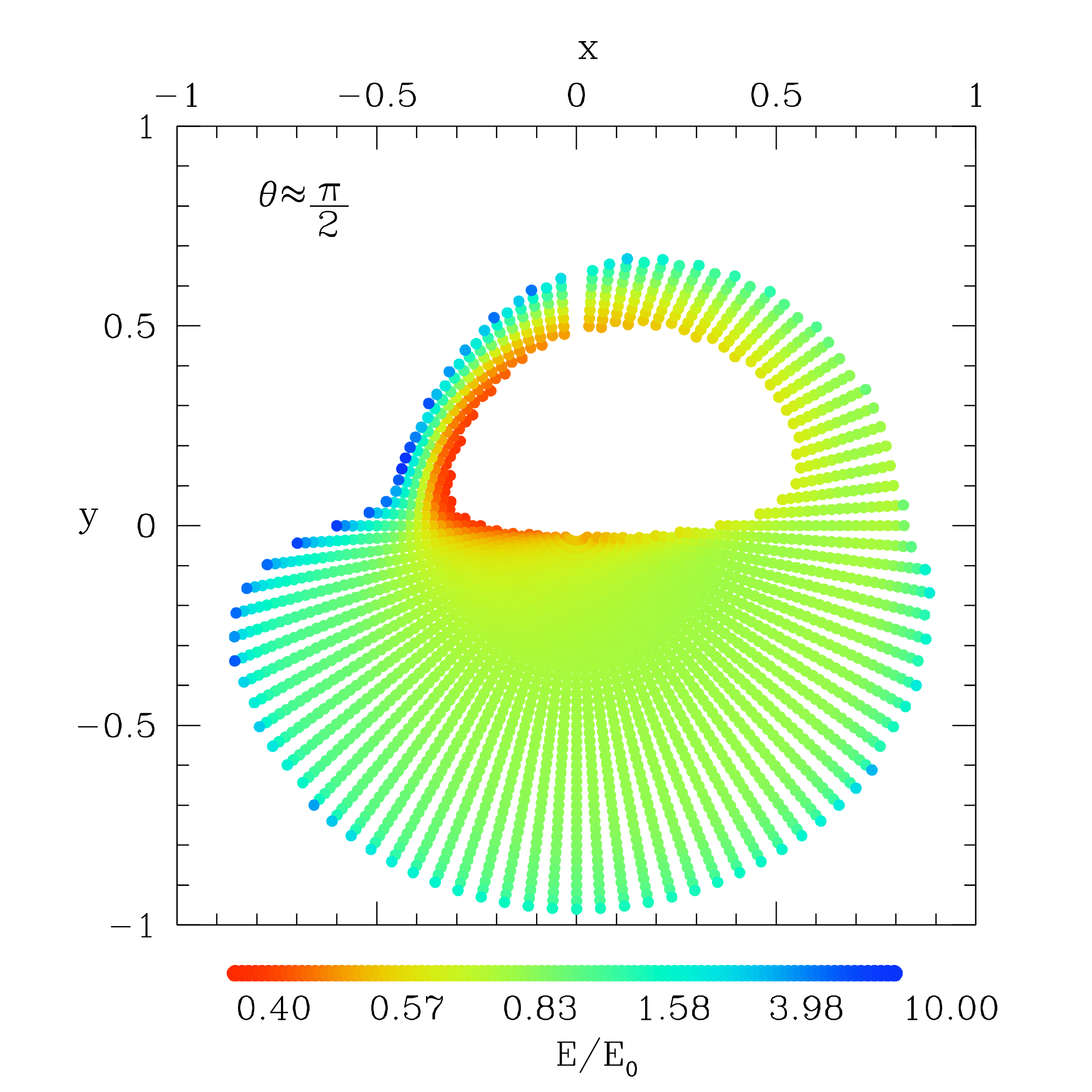}
\caption{
Projection of the sky for three different observers receiving neutrinos 
from a disc of radius $140r_h$.
The colour codes the ratio of the energy at reception to 
the energy at emission for neutrinos, as measured in the local ZAMO 
  frames 
(Section 2.4). All three observers are located at 3.1 horizon radii $r_h$ 
around a black hole of spin $a=0.95$. Their $\theta$ positions are 
approximately $0$, $\pi/4$ and $\pi/2$. The observer sky is parameterized 
with two angles $\alpha\in(0,\pi)$ and $\beta\in(0,2\pi)$ in the ZAMO frame.
The axis $\alpha=0$ points into the black hole and $\alpha=\pi$ points 
away from the black hole. The $x$ and $y$ coordinates in the figure are 
$(x,y)=(\frac{\alpha}{\pi} \sin\beta,\frac{\alpha}{\pi} \cos\beta)$. 
}
\label{fig:redshifts}
\end{center}
\end{figure}

Figure~\ref{fig:redshifts} shows the image of the accretion disc observed
from three locations near the black hole. Colour represents the redshift
(or blueshift) of neutrinos as they propagate from the emission point on 
the disc to the observation point.
The asymmetry of the images is caused by the rotation of the black hole.
%
%

\subsection{Local rates of $e^\pm$ creation}

\subsubsection{Reaction $\nu\bar\nu\rightarrow e^+e^-$}

The rate of $\nu\bar\nu$ annihilation at a given point depends only on
the local momentum distribution of $\nu$ and $\bar\nu$. 
We use the phase-space occupation number to describe this distribution;
it is denoted by $f$ for neutrinos and by $f^\prime$ for anti-neutrinos.
The local 
energy-momentum deposition rate is given by (see e.g. Birkl et. al. 2007),
\begin{eqnarray}
\label{eq:Qalpha}
  Q^{\alpha}_{\nu\bar\nu}=
   \int \frac{\textrm{d}^3p}{h^{3}}\frac{\textrm{d}^3p'}{h^{3}} 
   A
   (p,p') \left(p^\alpha+p^{\prime\alpha}\right)
       f(p)f'(p^\prime),
\end{eqnarray}
where 
\begin{eqnarray}
\label{eq:Aalpha}
  A
  (p,p')=\frac{\sigma_0 c^2}{4}
     \Big[ C_1 \frac{p^0p^{\prime 0}}{m_e^2c^2}(1-\cos\theta)^2 
	  +C_2 (1-\cos\theta)\Big],
\end{eqnarray}
where $p$ and $p^\prime$ are the four-momenta of neutrino and anti-neutrino,
$\theta$ is the angle between the spatial (3D) components of $p$ and 
$p^\prime$,
$\sigma_0=4m_e^2G_{\rm F}^2/\pi\hbar^4\approx 1.71\times 10^{-44}$cm$^{2}$, 
$G_{\rm F}$ is Fermi constant, $C_{1}=0.78$ and $C_{2}=1.06$. 
We neglect the mass of $\nu$ and $\bar\nu$, i.e. assume 
$p^\alpha p_\alpha=0$ and $p^{\prime\alpha} p^\prime_\alpha=0$.

\subsubsection{Reaction $\nu\to\nu e^+e^-$ in a strong magnetic field}

Neutrinos propagating in a strong magnetic field $B$ can create 
electron-positron plasma through the process $\nu\to\nu e^{+}e^{-}$ 
  (Kuznetsov \& Mikheev 1997; Gvozdev \& Ognev 2001).
A neutrino with four-momentum $p^{\alpha}$ and energy $E$ 
deposits energy-momentum into the $e^\pm$ plasma with the following 
rate 
\begin{eqnarray}
\label{eq:qnuB}
  q^\alpha = \frac{7 (c_v^2 + c_a^2)}{1728\pi^2}\, 
    \frac{\sigma_0}{\hbar}\,\left(eB\sin\psi\right)^2
   \ln{\left(\frac{\hbar p^0 eB\sin\psi}{m_e^3c^4}\right)} 
    \frac{p^0p^\alpha}{(m_ec)^2},
\end{eqnarray}
where $\psi$ is the angle between the magnetic field and the neutrino 
momentum, $c_{v}\approx0.96$ and $c_{a}=1/2$. 
The local energy-momentum deposition rate via 
  reactions $\nu\to\nu e^{+}e^{-}$ and $\nu^\prime\to\nu^\prime e^{+}e^{-}$ 
is given by
\begin{eqnarray}
\label{eq:QnuB}
   Q^{\alpha}_{\nu B}=\int \frac{\textrm{d}^{3}p}{h^{3}}
   \left[f(p)+f'(p)\right]\, q^\alpha(p).
\end{eqnarray}

\subsection{Integration of the energy deposition rate over volume}

The local energy-momentum deposition rates (eqs.~\ref{eq:Qalpha} 
and~\ref{eq:QnuB}) can be calculated in any frame of reference.
We chose the frame of Zero Angular Momentum Observer, ZAMO.
These observers do not move in $r$ or $\theta$ directions. Their
$\phi$ motion corresponds to zero angular momentum ($p_{\phi}=0$).
We use ZAMO frames because they are well defined everywhere outside the 
black-hole horizon (no observers can be static in Boyer-Lindquist 
coordinates inside the ergosphere). 
The local orthonormal coordinates of ZAMO, $d\tilde x^\alpha$, are 
related to $dx^\alpha=(dt,d\phi,dr,d\theta)$ by
\begin{equation}
  d\tilde x^{\alpha}=\mathcal{A}^\alpha_{~\beta}\,dx^\beta,
\end{equation}
where 
\begin{eqnarray}
\nonumber
  \mathcal{A}=\left( \begin{array}{cccc}
    \sqrt{-g_{tt}+(g_{t\phi})^{2}/g_{\phi\phi}} & 0 & 0 & 0\\
  g_{t\phi}/\sqrt{g_{\phi\phi}} & \sqrt{g_{\phi\phi}} & 0 & 0 \\
   0 & 0 & \sqrt{g_{rr}} & 0 \\
   0 & 0 & 0 & \sqrt{g_{\theta\theta}}  \\
\end{array} \right).
\end{eqnarray}
${\rm det}\mathcal{A}=\sqrt{-g}$ and the element of 
four-dimensional volume measured by ZAMO $d\tilde V$ is related to 
the Boyer-Lindquist coordinate volume $dV=dt\,d\phi\,dr\,d\theta$ by
\begin{equation}
   \dd \tilde V=\sqrt{-g}\,\dd t\,\dd\phi\,\dd r\,\dd\theta.
\end{equation} 
Let $\dd \tilde P^{\alpha}=Q^{\alpha}\dd \tilde V$ be four-momentum 
deposited in $\dd\tilde V$ as measured by ZAMO. 
The deposited energy measured by a distant observer is $\dd E =-c\,\dd P_t$. 
It can be expressed in terms of $\dd\tilde P^{\alpha}$,
\begin{equation}
  \frac{\dd E}{c} 
        =-g_{tt}\dd P^{t}-g_{t\phi}\dd P^{\phi}
        = \frac{\dd\tilde P^t}{\sqrt{-g^{tt}}}
         -\frac{g_{t\phi}}{\sqrt{g_{\phi\phi}}}\,\dd \tilde P^{\phi}, 
\end{equation}
where we used the transformation 
$\dd \tilde P^{\alpha}=\mathcal{A}_{~\beta}^{\alpha}dP^{\beta}$. 
This yields
\begin{equation}
  \dd E=\left(\frac{Q^t}{\sqrt{-g^{tt}}} -
     \frac{g_{t\phi}}{\sqrt{g_{\phi\phi}}}\,Q^\phi\right)
        \sqrt{-g}\,\dd t\,\dd\phi\,\dd r\,\dd\theta.
\end{equation}
The net energy deposition rate outside the horizon $dE/dt$ 
(measured by a distant observer) is given by
\begin{equation}\label{eq:Edotinf}
  \dot E = \int_{r>r_h} \left( \frac{Q^t}{\sqrt{-g^{tt}}} 
         -\frac{g_{t\phi}}{\sqrt{g_{\phi\phi}}} \,Q^\phi\right)\,
         \sqrt{-g}\,d\phi\,dr\,d\theta, 
\end{equation} 
where $r_h$ is the horizon radius.

\subsection{Numerical Method}

We compute the local rates of four-momentum deposition 
$Q^\alpha$ on a spatial grid. The problem is axially 
symmetric and the grid is set on the $(r,\theta)$ plane.
It covers the region 
  $r_h<r<r_{\rm max}$ 
and $0<\theta<\pi/2$ 
(we use the symmetry about the equatorial plane $\theta=\pi/2$).
  The radius $r_{\rm max}$ is chosen between $26r_h$ and $38r_h$, depending 
  on the black hole spin.
The grid has $25\times 20$ points. Its spacing is logarithmic in the 
$r$-direction and uniform in the $\theta$-direction.

We calculate $Q^{\alpha}$ according to equations~(\ref{eq:Qalpha})
and (\ref{eq:QnuB}). The distributions $f(p)$ and $f^\prime(p^\prime)$
are Lorentz-invariant (scalar) functions of four-momentum.
They remain constant along the neutrino trajectories (eq.~\ref{eq:Boltzman}).
Therefore, to calculate $f(p)$ measured by ZAMO at a given point,
it is sufficient to trace the neutrino with momentum $p$ back to its 
emission point, find its four-momentum there in the rest-frame of the disc, 
$p_{\rm em}$, and use the equality $f(p)=f_{\rm em}(p_{\rm em})$. The values 
of $f_{\rm em}(p_{\rm em})$ and $f_{\rm em}^\prime(p_{\rm em}^\prime)$ are 
provided by the disc models described in \S~2.1. 

For every point of the grid we trace back the neutrino trajectories
for 5000 directions uniformly distributed on the local ZAMO sky. 
Each direction is followed until the geodesic
reaches the disc, 
goes into the black-hole horizon or reaches a maximum radius that we set 
equal to $140r_h$. For trajectories coming from outside $140r_h$ or 
connecting to the horizon we set $f=f^\prime=0$. For trajectories connecting 
our grid point 
to
the disc, we find $f$ and $f^\prime$ as described above. 

The accuracy of our calculation of $Q^\alpha$ can be estimated by doubling 
the number of sampled geodesics; it is a few per cent. 
The accuracy of the volume-integrated quantity $\dot{E}$ is controlled mainly 
by the number of grid points. We checked it by repeating the same calculation 
with a coarser grid; the estimated error in $\dot{E}$ is smaller than 10 per 
cent (Table~1).

\begin{table}\label{table_accuracy}
\begin{center}
\begin{bf}{Code accuracy}\end{bf}
\begin{tabular}{|c|c|c|}
\hline
& Half spatial resolution & Twice geodesics\\
\hline
 a=0 & $|\Delta \dot E/\dot E|=0.03$ & 
       $|\Delta \dot E/\dot E|=5\times 10^{-3}$\\
\hline
 a=0.95 & $|\Delta \dot E/\dot E|=0.07$ & 
       $|\Delta \dot E/\dot E|=0.02$\\
\hline
\end{tabular}
\end{center}
\caption{Fractional change of the energy deposition rate when the grid
resolution is reduced from 500 to 240 points (left column), and when the 
number of geodesics is increased from $5\times10^{3}$ to $10^{4}$ 
(right column).}
\end{table}

\subsection{Comparison with previous works}

We compared our calculations with three previous works: 
Birkl et al. (2007), Asano \& Fukuyama (2001) and Popham et al. (1999). 

Birkl et al. (2007) calculated 
  $Q_{\nu\bar\nu}^\alpha$ and the corresponding volume-integrated 
  energy deposition rate $\dE$
outside the ergosphere for several
toy models of the neutrino source. They traced exactly the neutrino 
trajectories in the Kerr metric. Unfortunately, they incorrectly computed 
  $\dE$
by integrating $dP^t$ instead of $-dP_t$. For test purposes
we did a similar integration. We computed their models D and REF, which 
assume that neutrinos are emitted by an isothermal blackbody ring 
(see Table~1 in Birkl et al. 2007). 
The results agreed within the numerical errors of their and our calculations.

Asano \& Fukuyama (2001) calculated 
  $\dE$
on the rotational axis in Kerr 
metric as a function of $r$. The neutrino source was modeled as a blackbody 
disc with the temperature having a power-law dependence on radius. For test 
purposes, we repeated the calculation for their 
isothermal
disc model and obtained 
 $\dE(r)$ on the axis. The functional shape of $\dE(r)$
agrees with the shape 
of function $G(r)$ in Asano \& Fukuyama (2001) (see their eq.~18). The 
normalization of $G(r)$ appears to be arbitrary, so we were unable to 
compare the numerical values of 
  $\dE$.

Popham et al. (1999) made more realistic assumptions about the neutrino 
source, similar to our Model~C (Section~2.1). They evaluated numerically 
  $\dE$
for several values of accretion rate $\dM$ and black-hole spin $a$
(see their Table~3).
The neutrino geodesics in Kerr metric were replaced by straight lines.
When comparing their results with ours (which are presented in the next 
section) we found a significant disagreement, exceeding factor of 10.
Generally, we find that 
  $\dE$
of Popham et al. (1999) was overestimated. 
The dependence of 
  $\dE$
on $a$ that is suggested by Table~3 in Popham et al. (1999) is incorrect, 
apparently steeper than what we find numerically and estimate analytically 
(see next section).


\begin{figure}
\includegraphics[width=3.6in]{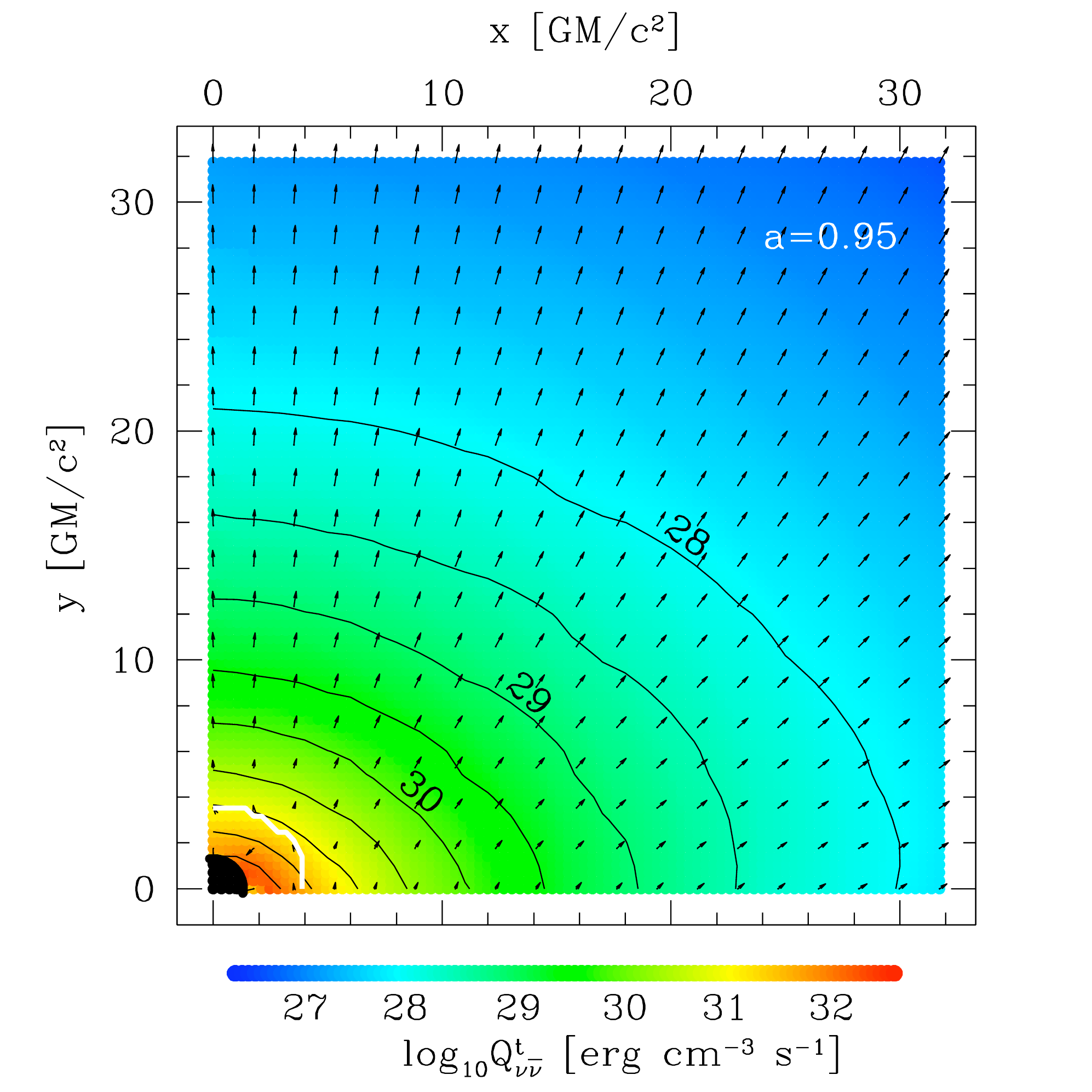}
\caption{Color-coded and contour plot for $\log_{10} Q^t_{\nu\bar\nu}$ 
around an accretion disc with $\dM=1M_{\odot}/s$. The black hole has 
mass $M=3M_\odot$ and spin parameter $a=0.95$. The horizon sphere (black) 
has the radius $r_h\approx 1.3GM/c^2$, and the inner edge of the disc 
(the marginally stable orbit) is at $\rms\approx r_g=2GM/c^2$. 
Arrows show the projection of $Q^{i}_{\nu\bar\nu}/Q^{t}_{\nu\bar\nu}$ 
on the plane of the figure. 
The white curve is where the radial component of 
  injected momentum $Q^r_{\nu\bar\nu}$
changes sign. 
  It roughly indicates 
the region where
the deposited energy may be lost into the black hole rather than escape
in an outflow.
  Disc Model~A with viscosity parameter $\alpha=0.1$ (Section 2.1) 
  was used in the calculation. Practically the same $Q^t_{\nu\bar\nu}$ is 
  found for Models~B and C, and for different $\alpha$ (e.g. $\alpha=0.01$).
}
\label{fig:Q95}
\end{figure}
%
\begin{figure}
\includegraphics[width=3.6in]{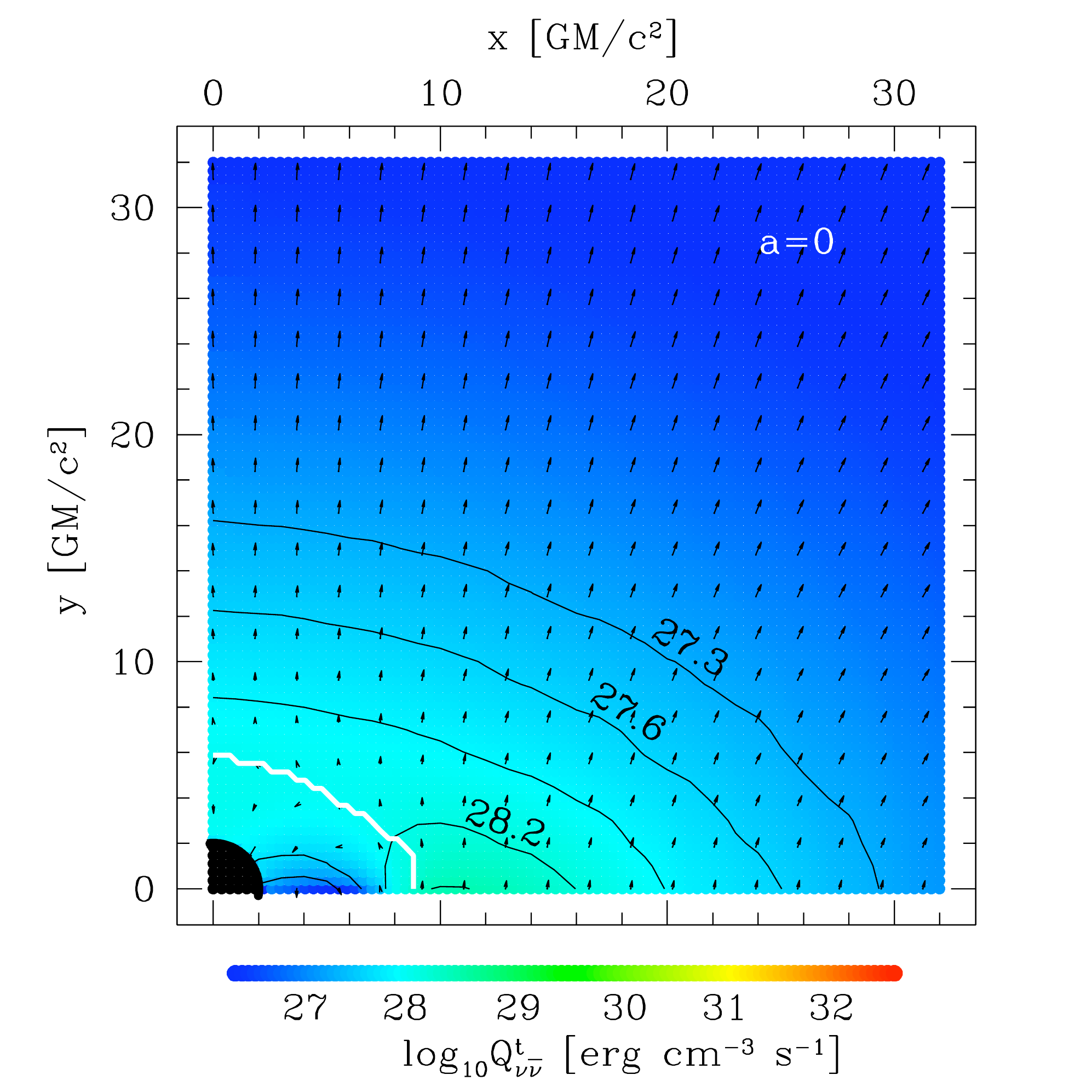}
\caption{Same as Fig.~\ref{fig:Q95} but for a non-rotating black hole,
$a=0$.}
\label{fig:Q00}
\end{figure}
%

\section{Results}
\label{sec:results}

The two processes of $e^\pm$ creation considered in this paper (Section~2.3) 
give the total energy deposition rate $\dot E=\dE+\dot{E}_{\nu B}$.
We first focus on $\dE$ and discuss the contribution
$\dot{E}_{\nu B}$ separately in Section~3.2.

\subsection{Energy deposition from $\nu\bar\nu\to e^+e^-$}

  Figures 2 and 3 show $Q^t_{\nu\bar\nu}$ for an accretion disc with 
  $\dM=1M_{\odot}$~s$^{-1}$ around a black hole of mass $M=3M_\odot$.
  In Figure~2 the black hole is assumed to be rapidly rotating ($a=0.95$),
  and in Figure~3 it is non-rotating ($a=0$). In the case of $a=0.95$,
  the deposition rate $Q^t_{\nu\bar\nu}$ peaks closer to the black hole 
  and reaches much higher values, because the disc extends to 
  smaller radii and emits a higher neutrino flux.

\begin{figure}
\begin{center}
\includegraphics[width=3.3in]{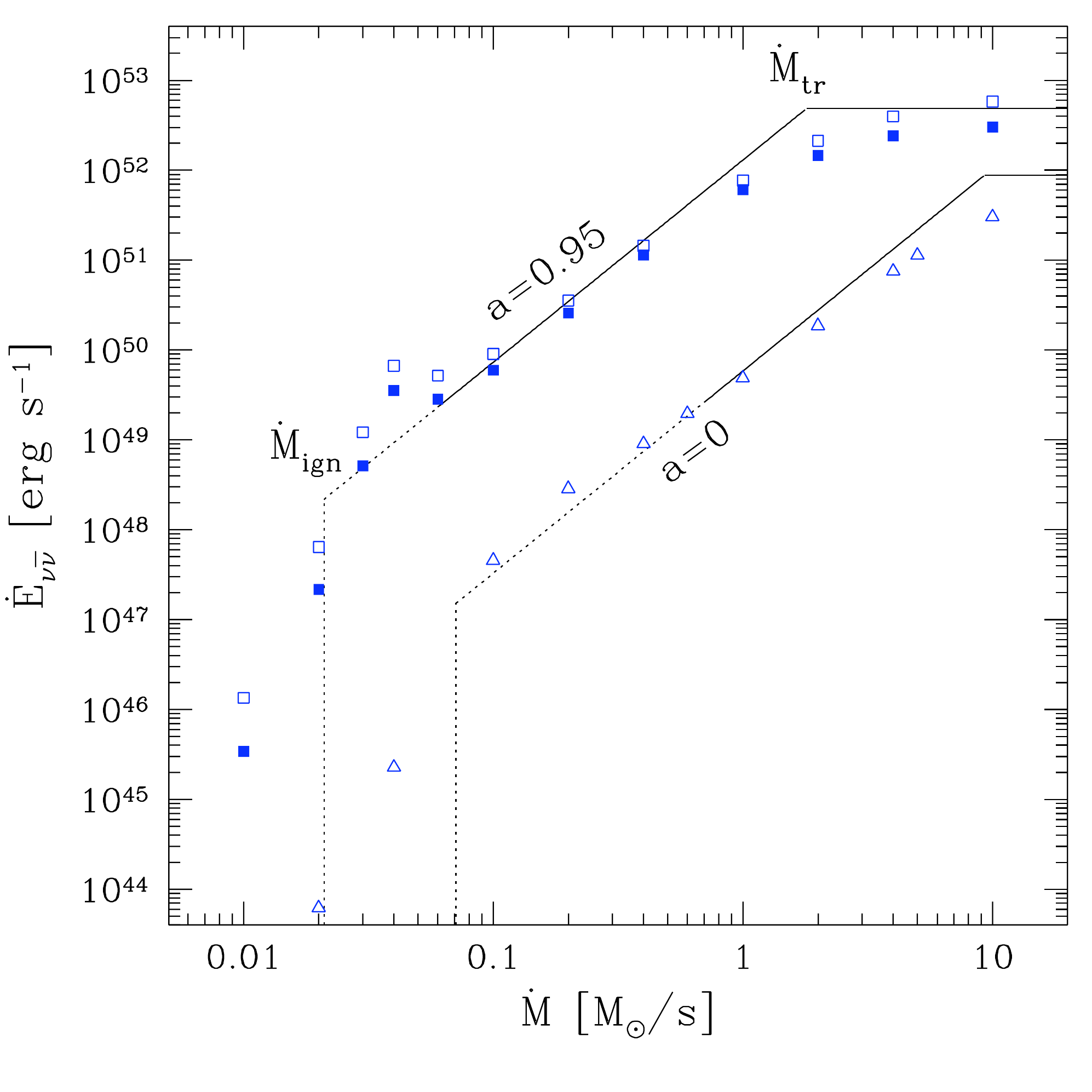}
\caption{$\dE$ as a function of $\dot M$ for a non-rotating black hole
($a=0$) and for a rotating black hole ($a=0.95$) of mass $M=3M_\odot$.
Open symbols show Model~A and filled symbols show Model~B (Section 2.1). 
The results of both models are well approximated by simple Model~C, 
which is shown in the figure by 
  lines;
the line is dotted at low $\dM$ 
where the disc is transparent to neutrinos.}
\label{fig:EdotVsMdot}
\end{center}
\end{figure}

Volume integration of $Q^t_{\nu\bar\nu}$ and $Q^\phi_{\nu\bar\nu}$,
as described by equation~(\ref{eq:Edotinf}), gives the 
  net
energy deposition 
rate due to $\nu\bar\nu$ annihilation outside the black-hole horizon, $\dE$.
Figure~\ref{fig:EdotVsMdot} shows $\dE$ as a function of the disc accretion 
rate $\dM$ for the two cases, $a=0$ and $a=0.95$. For all $\dM$,
$\dE$ is much higher when the black hole is rapidly rotating.

The results are 
  sensitive to $\dot{M}$ and $a$, but
remarkably insensitive to the details of the disc model.

(i) The uncertainty in the vertical
structure of the accretion disc leads to a small uncertainty in $\dE$
as illustrated by two extreme models described in Section~2.1: Model~A and 
Model~B. The results of both models are well approximated by simple Model~C 
(eq.~\ref{eq:C}).
Deviations of Model~C from Model~A arise mainly where Model~C does not 
accurately predict the neutrino flux from the disc, i.e.
where the disc is not strongly cooled by neutrino emission.

(ii) The uncertainty in viscosity parameter $\alpha\sim 0.1$ has 
almost no effect on $\dE$ as long as $\dMign<\dM<\dMtr$. In particular, 
Model~C in this range of $\dM$ is explicitly independent of $\alpha$. 
The two characteristic accretion rates $\dMign$ and $\dMtr$ 
depend on viscosity parameter $\alpha$ (see eq.~\ref{eq:pw}); 
in Figure~\ref{fig:EdotVsMdot} we assumed $\alpha=0.1$.

%
%

\begin{figure}
\begin{center}
\includegraphics[width=3.3in]{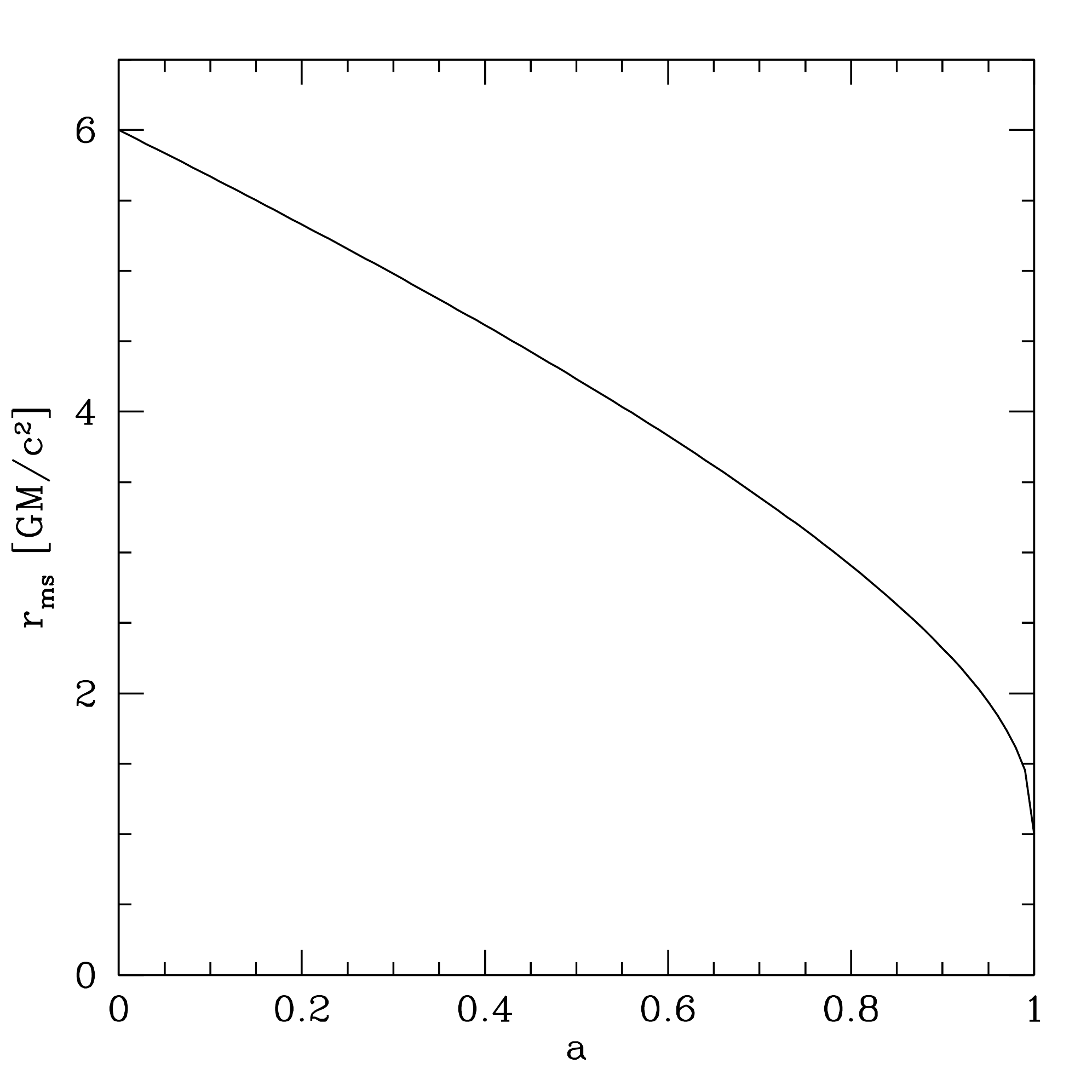}
\caption{Radius of the marginally stable orbit, $\rms$, as a function of 
the black-hole spin parameter $a$.} 
\label{fig:rmsvsa}
\end{center}
\end{figure}

\begin{figure}
\begin{center}
\includegraphics[width=3.3in]{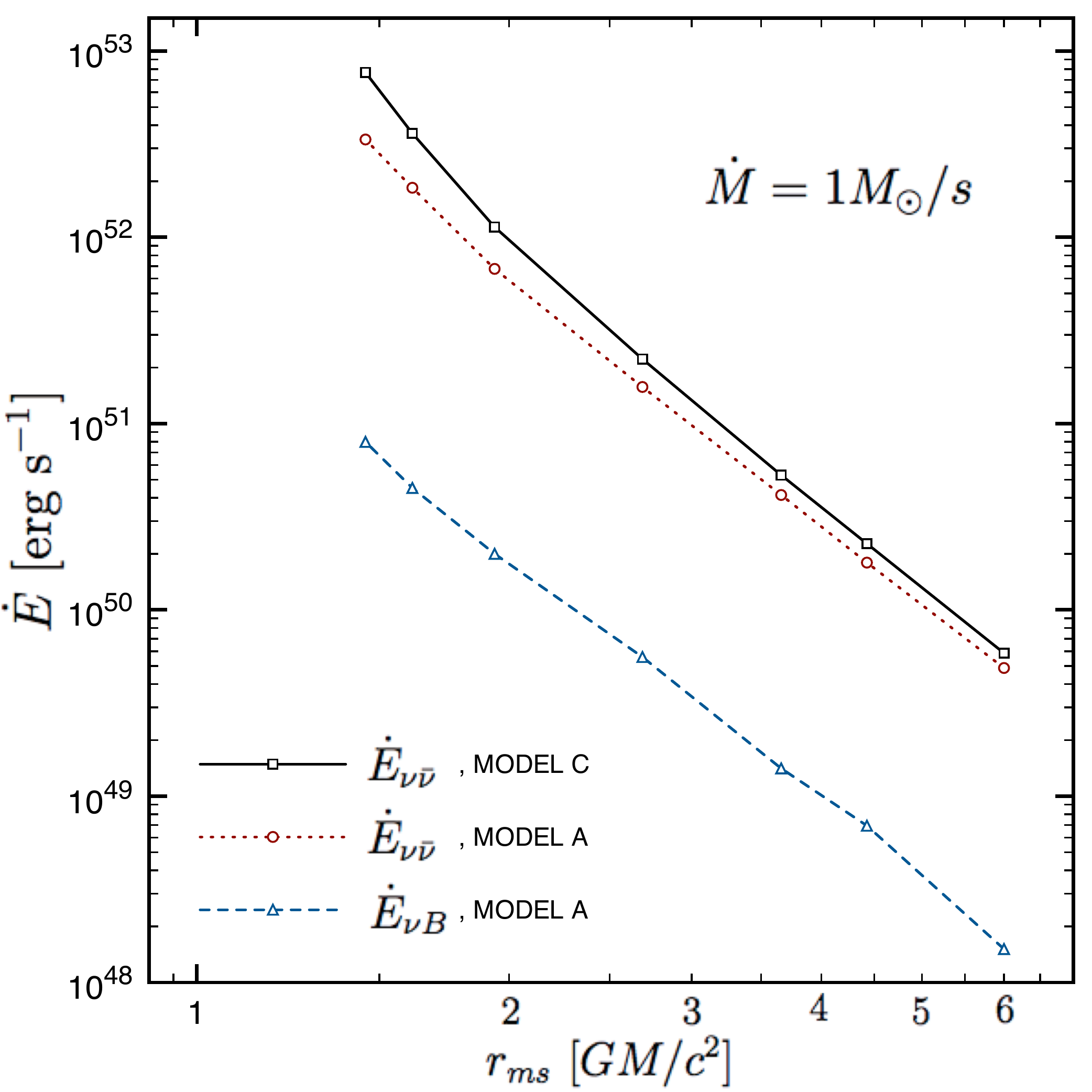}
\caption{$\dE$ and $\dot{E}_{\nu B}$ as functions of $\rms$ for fixed 
accretion rate, $\dM=1M_{\odot}/s$.}
\label{fig:EdotVsrms}
\end{center}
\end{figure}

The dependence of 
$\dE$
on the black hole spin, for a fixed $\dot M=1M_{\odot}/s$, is shown in 
Figure~\ref{fig:EdotVsrms}. Instead of using 
the spin parameter $a$ directly, it is more instructive
to plot $\dE$ versus 
radius of the last (marginally stable) orbit $\rms$. 
Then one can see the power-law dependence of $\dE$ on $\rms$: 
$\dE \propto \rms^{-4.8}$. This power-law is accurate for $\rms>r_g$
  which corresponds to 
$a<0.9$. For $\rms<r_g$, $\dE$ has a 
 somewhat stronger dependence on $\rms$.
The standard relation between $\rms$ and $a$ (e.g. Page \& Thorne 1974) 
is shown in Figure~\ref{fig:rmsvsa}. For non-rotating black holes 
$\rms=6 GM/c^{2}$ and for maximally rotating black holes $\rms=GM/c^{2}$.

%

\subsection{Energy deposition from $\nu\to\nu e^{+}e^{-}$ in a strong
magnetic field}

The four-momentum deposition due to reaction $\nu\to\nu e^{+}e^{-}$ 
is given by equation~(\ref{eq:QnuB}). It depends on the magnetic field.
We will assume the strongest field that could be expected,
\begin{eqnarray}\label{eq:EnuBdotproto}
   \frac{B^2}{8\pi}=P_{\rm max},
\end{eqnarray} 
where $P_{\rm max}$ is the maximum pressure in the disc; we find this 
pressure using the numerical disc model of 
  CB07.
For illustration, consider a toy model where the magnetic field is
perpendicular to the 
  disc,
uniform in the region $r<25r_h$ 
and zero outside this region.
The model overestimates a realistic magnetic field around an
accretion disc. 
We will show that even such a strong field gives a modest
$\dot{E}_{\nu B}$.
The maximum pressure $P_{\rm max}$ and $B=(8\pi P_{\rm max})^{1/2}$ 
depend on $\dM$ and $a$. The magnetic field ranges in our models from 
$\sim 6\times10^{13}$~G to $\sim8\times10^{15}$~G, increasing with $\dM$ 
and $a$.

Figure~\ref{fig:EdotVsrms} shows $\dot{E}_{\nu B}$ as a function of 
  $\rms$ for 
fixed $\dot M=1M_{\odot}/s$. 
It approximately follows the same scaling relation as we found for
$\dE(\rms)$, $\dot{E}_{\nu B}\propto r_{ms}^{-4.8}$.

\begin{figure}
\begin{center}
\includegraphics[width=3.3in]{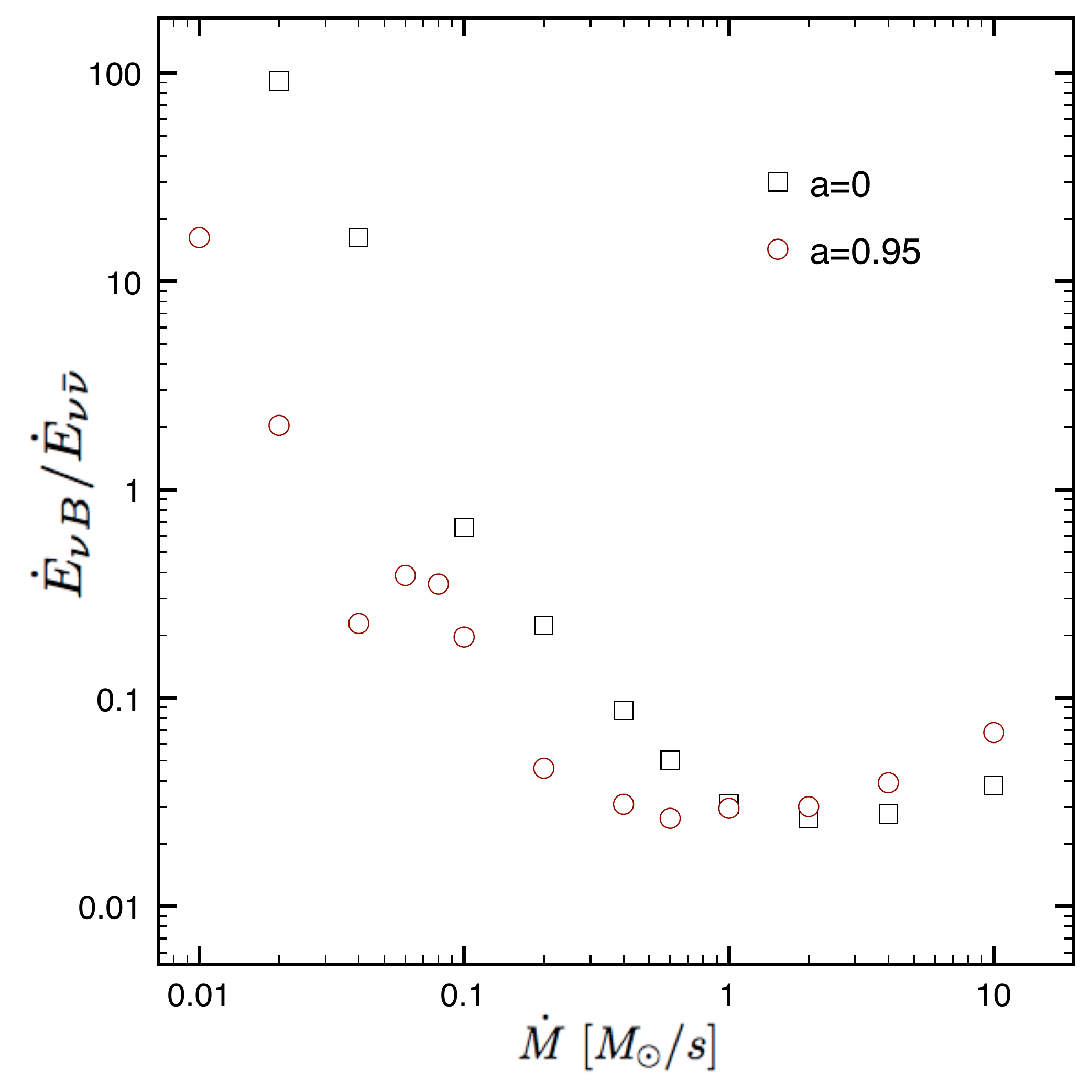}
\caption{Ratio $\dot E_{\nu B}$/$\dot E_{\nu\bar\nu}$ as a function of 
$\dot M$ for $a=0$ and $a=0.95$. The accretion disc is assumed to have  
viscosity parameter $\alpha=0.1$.}
\label{fig:EdotRatios}
\end{center}
\end{figure}
%
Figure~\ref{fig:EdotRatios} shows the ratio $\dot{E}_{\nu B}/\dE$ as 
a function of $\dM$. This ratio is small for accretion rates above $\dMign$
and varies slowly with $\dM$ or $a$. Thus, we find that $\dot{E}_{\nu B}$ 
makes a 
  small
contribution to the total $\dot{E}$.
This conclusion is valid for the most interesting range of accretion
rates $\dM>\dMign$. For $\dM<\dMign$, the neutrino flux quickly 
decreases, which strongly suppresses the $\nu\bar\nu$ annihilation above 
the disc. The reduction in reaction $\nu\to\nu e^+ e^-$ is less severe
and $\dot{E}_{\nu B}$ 
  exceeds
$\dE$. As a result, $\dot{E}_{\nu B}$ dominates the 
  energy deposition rate $\dot{E}$
when $\dM$ is below $\dMign$.

\begin{figure}
\begin{center}
\includegraphics[width=3.3in]{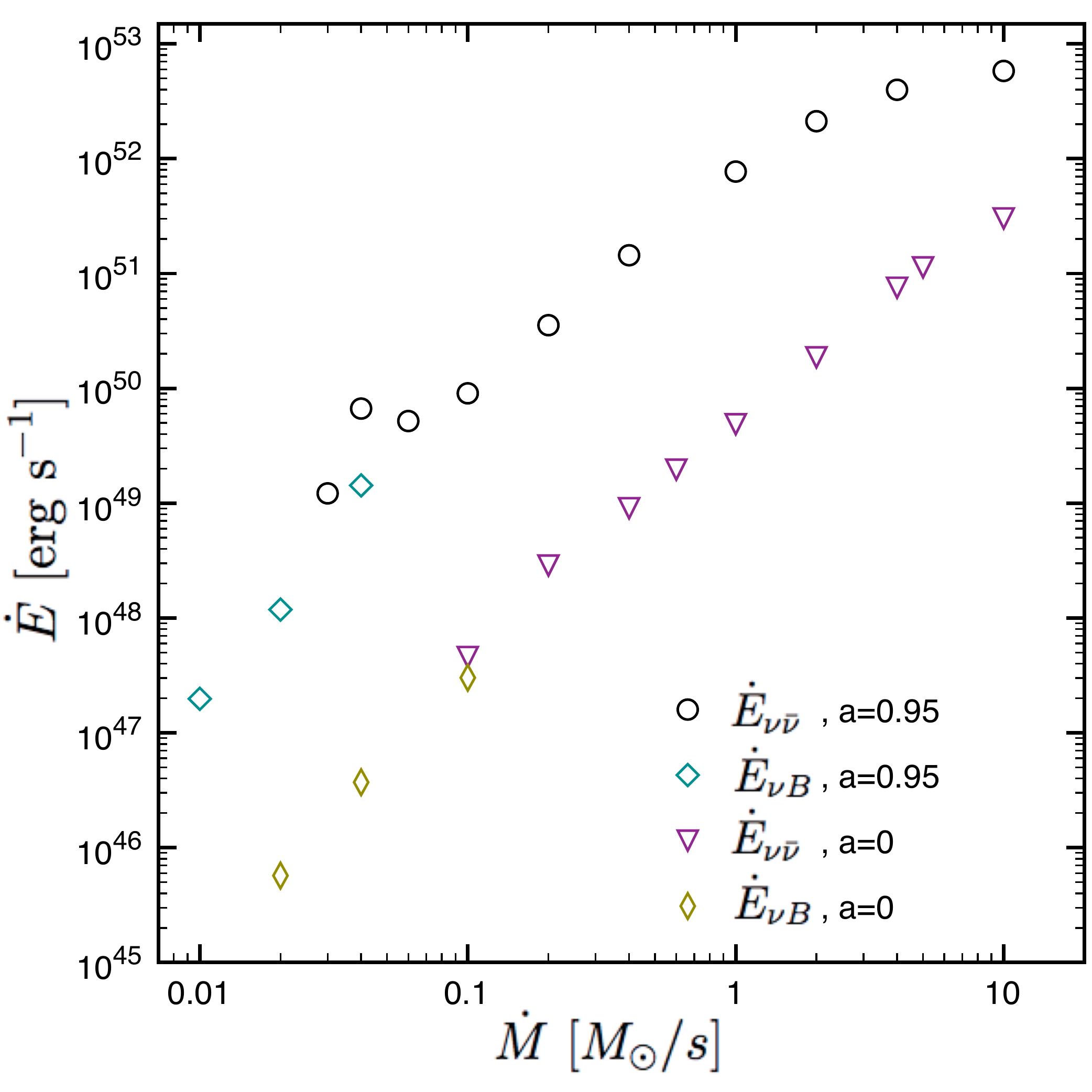}
\caption{Same as Fig.~\ref{fig:EdotVsMdot} but now including the contribution
of reaction $\nu\to\nu e^+ e^-$. This contribution 
  $\dot{E}_{\nu B}$
is shown only at 
$\dM<\dMign$, where it is important; 
  $\dot{E}_{\nu B}$
is small compared with $\dE$ at $\dM>\dMign$ (cf. Fig.~\ref{fig:EdotRatios}). 
}
\label{fig:EdotVsMdotMAG}
\end{center}
\end{figure}

  The dependence of the energy deposition rate on $\dM$ is summarized
in Figure~\ref{fig:EdotVsMdotMAG}, a modified
version of Figure~\ref{fig:EdotVsMdot}. It approximately represents 
the total 
  $\dot{E}$
by showing $\dE$ at $\dM>\dMign$ and $\dot{E}_{\nu B}$ at $\dM<\dMign$.
%
%

\subsection{Scaling of $\dE$ and $\dot{E}_{\nu B}$ with $\dM$, $M$ and $\rms$}

The $\dE$ dependence on 
  $\dM$, $M$ and $a$ 
may be estimated using simple
analytical arguments (Beloborodov 2008). The total rate of $\nu\bar\nu$
annihilation around the disc $\dN$ [s$^{-1}$] is proportional to the
volume of the main annihilation region $r^3$, which scales as $\rms^3$, 
and to the typical local annihilation rate in this region,  
$\dn\sim \sigma c n_\nu n_{\bar\nu}$.
Here $n_\nu$ and $n_{\bar\nu}$ are the number densities of neutrinos
and anti-neutrinos and $\sigma$ is the annihilation cross-section.
The cross-section scales with the energies of the annihilating 
$\nu$ and $\bar\nu$ approximately as $\sigma\propto E_\nu E_{\bar\nu}$,
which leads to 
\begin{equation}
  \dn\propto F_\nu F_{\bar\nu}.
\end{equation}
The energy fluxes $F_\nu$ and $F_{\bar\nu}$  
from an efficiently cooled disc ($\dMign<\dM<\dMtr$)
carry the released gravitational energy and do not depend on details of 
the disc structure. The fluxes scale with $M$, $\dM$, and $r$
as $F_\nu\sim F_{\bar\nu}\propto M\dM/r^3$. This gives
\begin{equation}
  \dN\propto r^3\dn\propto\frac{\dM^2}{\xms^3 M}.
\end{equation}
Here we assumed that the size of the main annihilation region $r$ is 
proportional to $\rms=\xms r_g$, where $\xms$ is determined by the 
black-hole spin parameter $a$ (see Fig.~5). 

The net heating rate $\dE$ may be estimated as 
$\dE\sim (E_\nu+E_{\bar\nu})\dN$, where
$E_{\bar\nu}\sim E_{\nu}\propto F_\nu^{1/4}$ is roughly estimated using
Model~C 
  (Section~2.1). 
This gives
\begin{equation}
  \dE\propto \rms^3F_\nu^{9/4}\propto \xms^{-15/4}\,\dM^{9/4} M^{-3/2}. 
\end{equation}
Our numerical results confirm the scaling $\dE\propto \dM^{9/4}$ and give a 
somewhat steeper dependence on $\xms$, $\dE\propto \xms^{-4.8}$ (Section~3.1).
The relation $\xms(a)$ can be substituted here to get the dependence 
$\dE(\dM,M,a)$.

  It is instructive to compare $\dot{E}_{\nu B}$ with $\dE$.
From equations~(\ref{eq:Qalpha}) and (\ref{eq:QnuB}) one can derive 
the following estimate,
\begin{equation}
   \frac{\dot{E}_{\nu B}}{\dE}\sim \frac{Q_{\nu B}^t}{Q_{\nu\bar\nu}^t}
    \sim 0.1\alpha_f\,\frac{U_B}{U_\nu}\,\ln{\left(\frac{B}{B_Q}\,E\right)},
\end{equation}
where $\alpha_f=e^2/\hbar c=1/137$, 
$B_Q=m_e^2c^3/e\hbar\approx 4.4\times 10^{13}$~G, $E$ is the average energy 
of neutrinos in units of $m_ec^2$, $U_\nu\sim F_\nu/c$ is the neutrino energy 
density and $U_B=B^2/8\pi$. One may expect that $U_B$ is proportional to 
the pressure in the disc $P$, which depends on viscosity parameter $\alpha$. 
Our numerical calculations gave $\dot{E}_{\nu B}/\dE\simlt 0.1$ for 
discs with $\alpha=0.1$ and $\dM>\dMign$ (Fig.~7).
Since $P\propto \alpha^{-1}$, models with smaller $\alpha$ would 
give a higher $\dot{E}_{\nu B}\propto\alpha^{-1}$. However, for discs
with $\dM>\dMign$, it will not dominate over $\dE$ in the plausible
range of $\alpha>0.01$.

For small accretion rates $\dM<\dMign$, $F_\nu$ is strongly suppressed and 
$\dE$ is suppressed as $F_\nu^{9/4}$. Since $\dot{E}_{\nu B}\propto F_\nu$ 
its suppression is less severe. As a result $\dot{E}_{\nu B}$ dominates over 
$\dE$ when $\dM<\dMign$ as we found numerically in Section~3.2.


\section{Conclusions}\label{sec:conclusions}

We have performed detailed numerical calculations of $e^\pm$ creation 
around hyper-accreting spinning black holes. We studied two reactions:
$\nu\bar\nu\rightarrow e^+e^-$ and $\nu\rightarrow \nu e^+e^-$ (in a 
strong magnetic field). The reaction of $\nu\bar\nu$ annihilation 
dominates the energy deposition rate around discs with $\dM>\dMign$, which 
are strong emitters of $\nu$ and $\bar\nu$.
We found that the net energy deposition rate due to this process, $\dE$,
is well approximated by a simple formula (see Fig.~4),
\begin{eqnarray}
\label{eq:Edot}\nonumber
      \dE & \approx & 1.1\times 10^{52}\,\xms^{-4.8}\,
               \left(\frac{M}{3M_\odot}\right)^{-3/2} \\
      & \times  & \left\{ \begin{array}{ll}
                    0                  &  \dM<\dMign \\
                \dot{m}^{9/4} \;\;     &  \dMign<\dM<\dMtr \\
          \dot{m}_{\rm trap}^{9/4}\;\; &  \dM>\dMtr \\
                           \end{array}
                  \right\}             \,{\rm erg~s}^{-1},
\end{eqnarray}
where $\dot{m}=\dM/M_{\odot}$~s$^{-1}$, $\xms=\rms(a)/r_{g}$, $r_g=2GM/c^2$,
and $\dMtr$, $\dMign$ are given in equation~(\ref{eq:pw}).
The dependence of $\dE$ on the black hole spin is huge: $\xms^{-4.8}$ 
varies by a factor of 200 for $0<a<0.95$. Note that $\alpha$ (viscosity 
parameter of the disc) enters the result only through $\dMign$ and $\dMtr$.
Our numerical simulations in this paper 
  are
limited to black holes with mass 
$M=3M_\odot$, and the $\dE$ dependence on $M$ is evaluated analytically.

The efficiency of $\nu\bar\nu$ annihilation can be defined as 
$\epsilon=\dE/L$ where $L$ is the total neutrino luminosity of the disc.
For example, $a=0.95$ 
  (which corresponds to $\xms=0.97$) 
gives $L\approx 0.15\dM c^2$ 
  (CB07)
and $\epsilon\approx 0.05\dot{m}^{5/4}$
for $\dMign<\dM <\dMtr$. Note that $\dE$ is defined in this paper as the 
total energy deposition rate outside the event horizon.
A fraction of the created $e^\pm$ plasma falls
into the black hole and does not contribute to the observed explosion 
(Fig.~\ref{fig:Q95}). The corresponding refinement of $\epsilon$
depends on the plasma dynamics outside the disc, which is affected by 
magnetic fields and hard to calculate without additional assumptions.

The obtained $\dE$ may be comparable to GRB luminosities $\Lobs$.
  A plausible typical
value for $\Lobs$ is $\sim 10^{51}$~erg/s
(it depends on the beaming angle of the observed explosion, which is 
usually hard to estimate from available data).
This power is easily supplied by neutrino heating if the black hole has 
a large spin, e.g. $a=0.95$, for a moderate accretion rate 
$\dM\simgt 0.3M_\odot$~s$^{-1}$. For a non-rotating black hole,
this mechanism of GRB explosion requires $\sim 10$ times higher accretion 
rates.


\label{lastpage}

\end{document}